\documentclass{article}

\PassOptionsToPackage{numbers, compress}{natbib}
\usepackage[preprint]{neurips_2022}

\usepackage[utf8]{inputenc} % allow utf-8 input
\usepackage[T1]{fontenc}    % use 8-bit T1 fonts
\usepackage{hyperref}       % hyperlinks
\usepackage{url}            % simple URL typesetting
\usepackage{booktabs}       % professional-quality tables
\usepackage{amsfonts}       % blackboard math symbols
\usepackage{nicefrac}       % compact symbols for 1/2, etc.
\usepackage{microtype}      % microtypography
\usepackage{xcolor}         % colors
\usepackage{graphicx}       %includegraphics
\usepackage{algorithm}
\usepackage{algpseudocode}

\title{Stronger is not better: Better Augmentations in Contrastive Learning for Medical Image Segmentation}

\author{
 Azeez Idris\thanks{Corresponding Author} \\
  Department of Computer Science\\
  Ames, IA 50011 \\
  \texttt{aidris@iastate.edu} \\
    \And
  Abdurahman Ali Mohammed\\
  Department of Computer Science\\
  Ames, IA 50011 \\
  \texttt{abdu@iastate.edu}
  \And
  Samuel Fanijo\\
  Department of Computer Science\\
  Ames, IA 50011 \\
  \texttt{sfanijo@iastate.edu}}

\begin{document}

\maketitle

\begin{abstract}
Self-supervised contrastive learning is among the recent representation learning methods that have shown performance gains in several downstream tasks including semantic segmentation. This paper evaluates strong data augmentation, one of the most important components for self-supervised contrastive learning's improved performance. Strong data augmentation involves applying the composition of multiple augmentation techniques on images. Surprisingly, we find that the existing data augmentations do not always improve performance for semantic segmentation for medical images. We experiment with other augmentations that provide improved performance. 
\end{abstract}

\section{Introduction}

Semantic segmentation is the task of classifying pixels into different classes. Accessing labeled medical data for supervised learning is costly both due to expensive experts and patient privacy issues. Thus, training deep learning models for semantic segmentation poses a huge challenge. 
%rephrase the reason behind and make it related to the first paragraph. (Why contrastive learning 
%annotations removed -> keep labels
% cross domain transfer learning training.
% Strong augmentations to be rephrased.
% remove we would like to use
% Introduce SimCLR is the most works are built on top of.
% Remove a lot of.
% Other works than SimCLR 
% False positive improved?

 Contrastive learning methods typically train a model by keeping similar images together while pushing dissimilar images apart in the embedding space. The reason behind the popularity of this approach is that it enables models to learn without human-annotated labels. This would save countless hours on labeling and classifying data.

Recent works have shown the usefulness of contrastive learning in several dense prediction tasks such as semantic segmentation. For example, SimCLR\cite{Chen2020ARepresentations} uses unlabeled images from the same distribution to train a semi-supervised learning system. DenseCL implements a self-supervised learning through the optimization of a pairwise similarity loss between two views of images at pixel levels \cite{Wang2021DensePre-Training}. Xie et al. also use pre-text pixel-level tasks to learn dense feature representations \cite{Xie2021PropagateLearningb}, Chen et al. improve the full construction of sample pairs through generation of self-contrastive foreground and background prototype of all pixels \cite{Chen2021SCNet:Prototypes}, while using pre-trained saliency detectors, Van Gansbeke et al. explore the use of object masks to cluster samples \cite{Gansbeke2021UnsupervisedProposals}. 

Since becoming viable for standard image classification tasks, self-supervised learning has continued to grow in popularity in the medical domain. For example, \cite{Zhou2020ComparingRepresentations,He2020Sample-EfficientScans,Liu2019AlignSupervision,Zhou2020ComparingRepresentations} are recent works which focus on application of contrastive learning to medical images, particularly in \cite{Zhou2020ComparingRepresentations}  where the authors proposed a Comparing to Learn (C2L) pre-training approach which learns from medical data with no manual annotations. There are also other works which focus on semi-supervised learning for medical data, for example, \cite{Cheplygina2018Not-so-supervised:Analysis,Wang2020FocalMix:Detection,Zhang2020ContrastiveText}

In this work, we investigate the effectiveness of self supervised contrastive learning methods based on one of its strongest components, strong data augmentations. We evaluate strong data augmentation using varying batch size, and pre-training weights.

\section{Experimental Setup}

For contrastive self-supervised learning, we use the SimCLR\cite{Chen2020ARepresentations} framework. The KVASIR-SEG\cite{Jha2019Kvasir-SEG:Dataset} dataset is used. It contains 1000 polyp images along with their ground truth annotations.The dataset split is 60/20/20 for training, validation and testing respectively. We utilize U-Net\cite{Ronneberger2015U-Net:Segmentation} for the downstream segmentation task.

We use the SimCLR approach for pretraining while only varying the data augmentation steps. For the fine-tuning phase on the other hand, we use Dice Loss as our loss function. We evaluate the performance of the down-stream semantic segmentation task using dice loss, IoU, FScore, Recall, and Precision.

%Hyperparameters to be added.

The strong data augmentation proposed in SimCLR\cite{Chen2020ARepresentations} are random cropping followed by resize back to the original size, random color distortions, and random Gaussian blur. Rather than these augmentations, we use basic augmentations including Resize, Rotate, and Horizontal Flip.

% We follow most of the experimental settings as in the Self-paced paper. The image size used is 256 * 256, while each experiment was run for 200 epochs. The batch-size is 16 while the starting learning rate is 0.0000001. The experiments conducted involves the Pre-training or the Fine tuning phase.

% \usepackage{booktabs}
% \usepackage{graphicx}
\begin{table}[t]
\centering
\caption{Performance comparison between SimCLR's strong augmentation versus simple augmentations based on different batch sizes and initialization weights. By using simpler augmentations rather than stronger augmentations, model performance improves for most cases. Numbers in bold indicate where basic augmentations outperform SimCLR's augmentations.}
\label{tab:my-table}
\resizebox{\columnwidth}{!}{%
\begin{tabular}{@{}clcccccccccccccccc@{}}
\toprule
 &
   &
  \multicolumn{1}{l}{} &
   &
  \multicolumn{2}{c}{\textbf{Dice Loss ↓}} &
  \textbf{} &
  \multicolumn{2}{c}{\textbf{IoU Score ↑}} &
  \textbf{} &
  \multicolumn{2}{c}{\textbf{FScore ↑}} &
  \textbf{} &
  \multicolumn{2}{c}{\textbf{Recall ↑}} &
  \textbf{} &
  \multicolumn{2}{c}{\textbf{Precision ↑}} \\ \midrule
Batch Size &
   &
  Weights &
   &
  SimCLR &
  Basic &
   &
  SimCLR &
  Basic &
   &
  SimCLR &
  Basic &
   &
  SimCLR &
  Basic &
   &
  SimCLR &
  Basic \\ \midrule
8 &
   &
  Random &
   &
  0.2300 &
  \textbf{0.1950} &
   &
  0.6928 &
  \textbf{0.7259} &
   &
  0.7699 &
  \textbf{0.8050} &
   &
  0.7967 &
  \textbf{0.8452} &
   &
  0.8282 &
  \textbf{0.8363} \\
16 &
   &
  Random &
   &
  0.1943 &
  \textbf{0.1686} &
   &
  0.7310 &
  \textbf{0.7571} &
   &
  0.8057 &
  \textbf{0.8314} &
   &
  0.8249 &
  \textbf{0.8537} &
   &
  0.8575 &
  \textbf{0.8693} \\
32 &
   &
  Random &
   &
  0.1690 &
  \textbf{0.1533} &
   &
  0.7562 &
  \textbf{0.7756} &
   &
  0.8309 &
  \textbf{0.8467} &
   &
  0.8516 &
  \textbf{0.8742} &
   &
  0.8711 &
  \textbf{0.8781} \\ \midrule
64 &
   &
  ImageNet &
  \multicolumn{1}{l}{} &
  0.1183 &
  \textbf{0.1154} &
  \multicolumn{1}{l}{} &
  0.8191 &
  \textbf{0.8202} &
  \multicolumn{1}{l}{} &
  0.8817 &
  \textbf{0.8846} &
  \multicolumn{1}{l}{} &
  0.8992 &
  \textbf{0.9015} &
  \multicolumn{1}{l}{} &
  0.9033 &
  0.9031 \\ \bottomrule
\end{tabular}%
}
\end{table}
\section{Results}
We begin with random weight initialization at the pre-training phase. Based on SimCLR\cite{Chen2020ARepresentations}, increasing the batch size improves model performance. Thus, we increased the batch size from 8, 16, and finally to 32 for the randomly initialization weights. Using basic augmentations improve model performance consistently for IoU between 2 to 3 percent in each case. The improvement is also consistent for the FScore, recall, and precision. 

To evaluate SimCLR\cite{Chen2020ARepresentations} even better, we increased the batch size to 64, and used ImageNet pretrained weights. The basic augmentations still do not perform worse than SimCLR\cite{Chen2020ARepresentations} for most of the metrics.

\section{Discussion and conclusion}
Although the increase in batch size improves model performance, using a stronger augmentation does not improve performance. We hypothesize that the augmentations needed to improve contrastive self-supervised learning need not be strong, but must be relevant and useful for the given dataset. In this case, strong augmentations used to train ImageNet are not better for training medical semantic segmentation datasets.

Future work would consist of determining the impact of other components of SimCLR while evaluating its performance for medical datasets. And also providing theoretical and logical observations of why these components fail when they do.

\bibliography{references}

\end{document}